# *Cosmic Inflation, the Beginning of the Universe, and the Origin of the Various Elements*

*by*

*Roger Ellman*

Abstract

The cosmic inflation hypothesis, its relation to fundamental theory on the beginning of the universe, and the light that both shed on how the various elements and their relative amounts came into existence are addressed and analyzed.

The fundamental factors controlling the origin of the universe *ex nihilo* are developed and from them the following alternative hypotheses are developed and presented:

- an alternative to the inflation hypothesis, the purpose of which hypothesis is to explain how the structure in the universe developed from the initially unstructured and symmetrical Big Bang, and

- an alternative to the hypothesis that the various elements in the universe were formed by nuclear fusion in stars [not denying that it may occur occasionally].

Roger Ellman, The-Origin Foundation, Inc.
320 Gemma Circle, Santa Rosa, CA 95404, USA
RogerEllman@The-Origin.org
http://www.The-Origin.org

## Cosmic Inflation, the Beginning of the Universe, and the Origin of the Various Elements

by

*Roger Ellman*

Fermilab's in-house magazine of February 2008 ran a spoof personal ad that stated: "mature paradigm with firm observational support seeks a fundamental theory in which to be embedded". It was referring to inflation – a period of exponential expansion of the universe thought to have taken place about $10^{-35}$ seconds after the big bang and which, although able to account for the large-scale appearance of the universe, lacks a firm theoretical footing. [1]

[The problem that the inflation hypothesis attempts to address is that of accounting for that the present universe is not perfectly symmetrical and homogeneous yet its beginning at the very first instant of the beginning, the Big Bang, had to be perfectly symmetrical and homogeneous. The inflation hypothesis posits initial expansion of the universe immediately after its beginning at a rate far exceeding the speed of light so that various parts of the universe became out of communication with the rest and consequently the overall symmetrical and homogeneous nature could not maintain.]

The problem of a theoretical footing for the inflation hypothesis is directly related to another fundamental cosmological problem, the very beginning of the universe from nothing, *ex nihilo*.

- The universe is thought to have had to have started at a singularity, a dimensionless, volumeless point. That is required because it is thought that any other start involves an infinite rate of change from the immediately prior state of no universe to what followed.

- But, how could the universe have started at a singularity, a point, which is dimensionless, volumeless, and not conceivably able to deliver anything, not able to be the origin of anything ?

- And, how could the universe come into existence without a violation of conservation, without "getting something from nothing" ?

### *The Original "Singularity"*

It turns out that the solution to the problem of the very beginning of the universe is also the solution to the problem of inflation. The function of the original singularity is to avoid an initial infinite rate of change at the beginning of the universe. That objective does not require a dimensionless point. It only requires that the <u>form</u> of the beginning must be such that all of its mathematical derivatives, that is rates of change, be finite. The only form that does that and matches to the initial value of `zero` and does not increase without limit is the `[1 - Cosine]` function, which in its `half cycle` from `zero angle` to `180°` [`π radians`] performs a smooth transition from zero amplitude to its maximum.

In order to not violate conservation that form and its opposite must have simultaneously arisen as in the form `±[1 - Cosine]` so that they balance out to nothing, the pre-origin *nihilo*.

An <u>object of non-zero volume</u> oscillating in that fashion avoids infinite rate of change. For the purpose of being the origin of the universe it is a "singularity" and is one that can deliver a universe which a dimensionless point cannot.

### *The Origin of the "Big Bang" and the Solution to "Inflation"*

The origin of the universe was not a dimensionless singularity, a true volumeless point. It was a volume of radius equal to about $4 \cdot 10^7$ *meters*.[3] That resolves the inflation problem. The universe did not have to extremely rapidly and unreasonably [far exceeding the speed of light] expand a super-brief moment after its beginning. It began already so inflated.

[Why that particular size ? Why any size ? There was nothing to compare it with. It simply was. The value is derived and calculated[3] from the estimated number of particles in the universe[7]. See the derivations associated with references[5 & 3].]

Examining further the original non-zero volume "singularity" the change from nothing to something, both itself and its consequences, were subject to the restriction of the impossibility of a material infinity as well as the requirement of conservation. Resulting further implications are:

- the rate of change was finite; that is, rather than an instantaneous jump from nothing to something there had to have been a gradual transition at a finite rate of change;

- the rate of change of the rate of change (in calculus the 2nd derivative) was also finite; that is, rather than an instantaneous jump from zero to some non-zero rate of change there had to have been a gradual transition;

- similarly for the rate of change of the rate of change of the rate of change (in calculus the 3rd derivative) and so on *ad infinitum*.

These require that the change took place in a manner describable either as a natural exponential or some form of sinusoid. Only those forms can assure there being no derivative that is infinite as can be seen by examination of their expansions into infinite series.[4] There are five forms of that type; the: sine, cosine, hyperbolic sine, hyperbolic cosine, and natural exponential.

In addition to the finite derivatives requirement there are other requirements to match such a function to the real situation: the function must not be open-ended, that is it cannot ever have an infinite amplitude, and the function must smoothly match the *0* condition at *t = 0*. The only function that meets all of the requirements is the *U(t)* presented in the figure and equation below.[4]

In the figure, the equation which follows it, and the discussion, $t_0$ is the instant of the beginning and *U(t)* is the quantity (the substance of the <u>U</u>niverse) that changes with time, *t*.

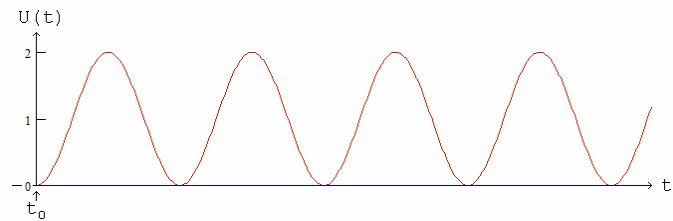

*Figure 1*
*Form For The Beginning of the Universe*

The curve of the above is that *U(t)* is of the form:

(1)     U(t) = 1 − Cos(2πft)]         t ≥ $t_0$
        U(t) = 0                      t < $t_0$.

This form for `U(t)` has the uniform content of the `4·10⁷ meters` radius volume starting at being zero throughout and smoothly changing with all rates of change finite. Thus, there is no infinite change from nothing to something.

Furthermore, at every instant the substance of `U(t)` is the same, uniform, throughout that initial volume. At every instant there is no difference anywhere within that volume. There can be no mensuration, no dimensioning. It is all the same place everywhere within that volume. Thus it is a *singularity*.

### *The Origin of the Various Elements*

Physicists' current hypothesis as to the origin of the various elements is that the heavier elements [those requiring added energy to be formed by fusion vs. those lighter elements that release energy in fusion] were formed in stars [the only place that could conceivably provide the necessary conditions for heavier element fusion]. Theory of such heavier element formation in stars requires that the star have evolved for at least about 700 million years before heavier element production could occur.

Recently heavier elements [iron in particular] have been detected in stars only about 900 million years old [per measurements of redshift]. That would leave only about 200 million years for the earliest stars to have formed – seemingly much too little time for the collection and gravitational "condensation" of enough hydrogen for star formation and the launching of their hydrogen fusion reactions, especially as compared to the 700 million year evolution time for the star to be able to form heavier elements.

The hypothesis of the heavier elements having been formed in stars has become largely untenable, its principal remaining support being the lack of an alternative hypothesis.

With regard to the lighter elements that do not require added energy to be formed by fusion, but release energy in fusion, Nikos Prantzos of the Institut d'Astrophysique de Paris in his article *Origin and Evolution of the Light Nuclides*, [6] states as follows,

> "In their monumental study on *Synthesis of the Elements in Stars*, Burbidge et al. (1957, B²FH) recognized the difficulty of finding a nuclear process able to synthesize the light nuclides D, $^6$Li and $^7$Li, $^9$Be, $^{10}$B and $^{11}$B. Indeed, these nuclides are so fragile (as revealed by their binding energies …) that they are consumed in stellar interiors, once hydrogen-rich material is brought to temperatures [sufficiently high] … ."

Prantzos concludes,

> "The *x-process* (as they called the unknown nucleosynthetic mechanism) turned out to be the most complex of all the nucleosynthetic processes envisioned in B²FH. Despite 50 years of progress in theory and observation, it is still unknown where most of $^3$He and $^7$Li and a large fraction of $^{11}$B come from. The origin of early $^6$Li remains equally mysterious, while the degree of astration of D in the solar neighborhood is poorly known."

Thus, as with the heavier elements, the formation of the lighter elements is only poorly supported by theory.

Another difficulty with the theory of the formation of the heavier elements by fusion in stars is the problem of the neutron. While a free neutron readily decays into a proton plus an electron and a neutrino, the inverse reaction generally does not operate. That is, an attempted combination of an electron plus a proton and energy results not in a neutron but in a Hydrogen atom. The electron takes up an orbit around the proton rather than merging with it.

That behaviour makes it difficult for heavier elements, which contain a number of neutrons in their nucleus, to form by fusion in stars.

An alternative theory that largely renders element formation in stars irrelevant [while not denying that it may occur occasionally] begins from the point of view that all of the various elements are decay products of a primal / original radioactive decay of an immense original nuclear-type structure, an immense nuclear form, charge-neutral and containing all of the mass of the universe. In other words, that the various elements came about via <u>fission</u> of an original "cosmic egg" nuclear type rather than via <u>fusion</u> of individual original fundamental particles.

With regard to the issue of inflation it was stated above,

> "The origin of the universe was not a dimensionless singularity, a true volumeless point. It was a volume of radius equal to about $4 \cdot 10^7$ *meters* as derived and calculated from the estimated number of particles in the universe[7]."

Again, in other words, the origin was a single immense "particle" that explosively decayed into the myriad particles of the resulting universe. To develop that concept requires returning to equation *(1)* the conceptual form of the beginning of the Big Bang.

There are two problems with equation *(1)* as the form of the Origin. The first is that an infinite rate of change still remains; the envelope of the oscillation has an infinite rate of change at $t=t_0$. Viewed in a mathematical or graphical sense without any consideration of the physical reality represented, the envelope discontinuity at $t=t_0$ is not a difficulty. The only quantity that actually exists and is varying is the overall $U(t)$. The envelope is merely our perception of a characteristic of the wave form. The actual varying quantity has no discontinuity at $t=t_0$. However, looking at the situation in a physical sense rather than purely mathematically, the beginning of the non-zero $U(t)$ is the beginning of the matter and energy of the universe and is not insignificant.

That infinite rate of change in the envelope at $t=t_0$ is no more acceptable than was the infinite rate of change encountered in the original analysis of the probable beginning and it must be corrected by the same kind of reasoning as then pursued: the envelope, also, had to originate as a *[1 - Cosine]* form of oscillation, which is the only form that avoids an infinite rate of change and matches the requirements of the situation.

The second problem with equation *(2)* is that the positive $U(t)$ so far addressed had to be precisely accompanied by an equal but opposite negative $U(t)$ so that conservation is maintained; the sums, the totality, before and after time $t_0$ are identical. There can be no "something from nothing".

That original envelope oscillation was at a lesser frequency than original wave by the definition of a wave form envelope. If it were at a greater frequency then the roles (envelope and wave) would be reversed. If it were at the same frequency it would not act as an envelope and the infinity problem would remain. Designating the envelope frequency $f_{env}$ and the frequency of the wave oscillation within the envelope $f_{wve}$ then the envelope would be of the following form.

*(2)*     $U_{env} = [1 - Cos(2\pi \cdot f_{env} \cdot t)]$

The wave is, as before, of the form of equation *(1)*.

*(3)*     $U_{wve} = [1 - Cos(2\pi \cdot f_{wve} \cdot t)]$

The combination of the envelope modulating the wave is then of the form

*(4)*     $U(t) = \pm [U_{env}] \cdot [U_{wve}]$
              $= \pm [1 - Cos(2\pi \cdot f_{env} \cdot t)] \cdot [1 - Cos(2\pi \cdot f_{wve} \cdot t)]$

where the $\pm$ is to account for the "negative" $U(t)$, the equal but opposite oscillation needed to maintain conservation with the "positive" $U(t)$.

However, the form of $U(t)$ of equation *(4)* still does not resolve the problem of an infinite rate of change at $t_0$. The *[1 − Cosine]* envelope itself begins at $t_0$ with a sudden step from zero to its full amplitude. Therefore, it is again necessary to introduce yet another envelope of *[1 − Cosine]* form to prevent the infinite rate of change at $t_0$ in the prior envelope. That correction will in turn require still another such correction and so *ad infinitum.* The $U(t)$ resulting from that is then as equation *(5)*.

$$(5) \quad U(t) = \pm \prod_{i=1}^{i=\infty} \left[1 - \cos(2\pi \cdot f_{env_i} \cdot t)\right] \cdots \cdot \left[1 - \cos(2\pi \cdot f_{wve} \cdot t)\right]$$

where the $\prod$ symbol (a large $\pi$) means the <u>p</u>roduct of the indicated factors.

[For reasons developed in reference [3] the effective number of those envelopes was not infinite, but rather was limited by two effects. One was a bandwidth type of limitation analogous to electronic bandwidth and related to the increasing frequencies in the expansion of equation *(5)* as $i$ increases. That is developed in reference [5]. The second is a mathematical effect in the expansion of equation *(5)* that is developed in reference [3]. The number of those envelopes is causally the number of particles [7] produced in the original Big Bang as developed in [3]].

That $U(t)$ was a pair of extremely complex oscillations. It gave birth to all of the particles of the universe, and as such was itself a pair of immense, complex, charge-neutral of necessity, super-particles. As the neutron is the only charge-neutral atomic particle, the pair of super particles was like an immense super-neutron and its opposite, its anti-particle. That first instant of the universe, the starting of the pair of oscillations, $\pm U(t)$, was, the moment that they started, the starting of the existence of a pair of complex gigantic atomic nuclei, composed solely of neutrons, that, when the two are taken together, contained all of the mass / energy of the universe in their immense mass.

Each was unstable, of course; it was the most unstable nuclear structure that could be. They immediately decayed in an immense explosion of energy and particles, the event now called the "Big Bang". Their complexity, which gave us the about myriad particles of our universe, resulted from the myriad successive envelopes, all existing and acting simultaneously from the beginning of course. The envelopes themselves were essential in order to avoid an infinite rate of change, as already presented above.

Equation *(5)* involves the expression *[1 − Cos(2π·$f_{env}$·t)]* multiplied by itself for $i = \infty$ number of times. A plot of that expression exponentiated, *[1 − Cos(2π·$f_{env}$·t)]$^n$*, appears in Figure 2, below, for several values of the exponent, $n$. From the plot it is clear that the central peak progressively narrows as $n$ increases.

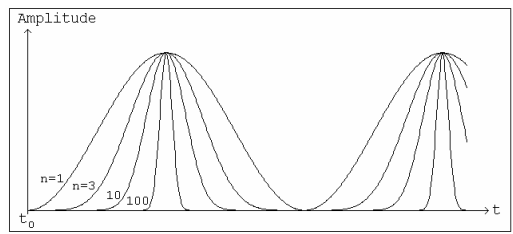

*Figure 2, [1− Cos(2π·$f_{env}$·t)]$^n$*

When $n$ becomes very large the plot appears as in Figure 3, below.

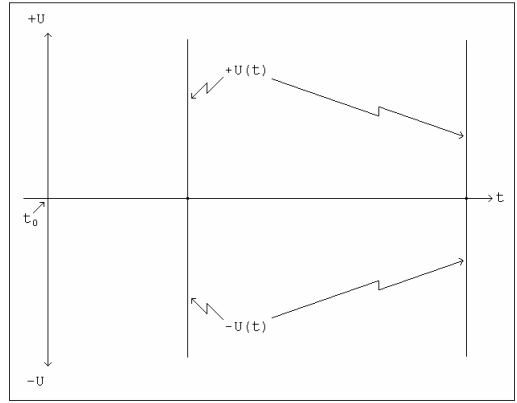

*Figure 3, $[1- Cos(2\pi \cdot f_{env} \cdot t)]^{\infty}$*

Because it was so extremely unstable, the explosive radioactive decay of $U(t)$ began instantly upon the oscillation's beginning at time $t_0$ the so immediate decay undoubtedly occurred after only a minute, an infinitesimal, portion, of the very first cycle had passed. It had to have been long before the first "spike". In that sense the initial event was very small, tenuous, hardly more than nothing because the instantaneous amplitude of $U(t)$ at that moment (the height of the curve above zero at that moment long before the first "spike") was also infinitesimal. It was hardly more than, essentially zero.

In that sense, the way that the universe started at all becomes a little more comprehensible. There was essentially almost no difference between "nothing", on-going absolute nothing, and the first infinitesimal moment of the original $U(t)$, the original oscillation.

Yet, it contained the entire universe.

Most probably the extreme instability and consequent immediate explosive decay account for the survival of the universe beyond its first moment, for otherwise the equal and opposite initial oscillations should have mutually annihilated.

### *The Origin of Structure in the Present Universe*

Several aspects caused the original Big Bang explosion to fail to be perfectly spherically symmetrical, resulting in the universe's varied structure:

- The explosion and projection outward was of particles, not a smooth continuous substance and no arrangement of particles could be perfectly uniform and symmetrical;

- The explosion was analogous to a radioactive decay of a very large and complex nuclear type proceeding in stages of decay down to the ultimate end of the chain; and

- The entire original volume contents could not perfectly simultaneously explode or decay so that the progress of the process through its various decay chains although extremely rapid defeated uniformity and symmetry.